\documentclass[pmlr]{jmlr}% new name PMLR (Proceedings of Machine Learning Research)

\usepackage{longtable}% for long tables

\usepackage{booktabs}
\usepackage{algorithmic}
\usepackage{caption}
\usepackage{comment}
\usepackage[load-configurations=version-1]{siunitx} % newer version
\theorembodyfont{\upshape}
\theoremheaderfont{\scshape}
\theorempostheader{:}
\theoremsep{\newline}

\jmlrvolume{204}
\jmlryear{2023}
\jmlrworkshop{Conformal and Probabilistic Prediction with Applications}
\jmlrproceedings{PMLR}{Proceedings of Machine Learning Research}

\title[Enterprise Disk Drive Scrubbing Based on Mondrian Conformal Predictors]{Enterprise Disk Drive Scrubbing Based on Mondrian Conformal Predictors }

 \author{\Name{Rahul Vishwakarma} \Email{rahuldeo.vishwakarma01@student.csullb.edu}\\
 \addr California State University Long Beach, 1250 Bellflower Blvd, Long Beach, CA 90840, United States
 \AND
 \Name{Jinha Hwang} \Email{jinha.hwang01@student.csulb.edu}\\
 \addr California State University Long Beach, 1250 Bellflower Blvd, Long Beach, CA 90840, United States
 \AND
 \Name{Soundouss Messoudi} \Email{soundouss.messoudi@hds.utc.fr}\\
 \addr HEUDIASYC - UMR CNRS 7253, Universit\'e de Technologie de Compi\`egne, 57 avenue de Landshut, 60203 Compi\`egne Cedex - France
 \AND
 \Name{Ava Hedayatipour} \Email{ava.hedayatipour@csulb.edu}\\
 \addr California State University Long Beach, 1250 Bellflower Blvd, Long Beach, CA 90840, United States
 }

\editors{Harris Papadopoulos, Khuong An Nguyen, Henrik Boström and Lars Carlsson}

\begin{document}

\maketitle

\begin{abstract}
Disk scrubbing is a process aimed at resolving read errors on disks by reading data from the disk. However, scrubbing the entire storage array at once can adversely impact system performance, particularly during periods of high input/output operations. Additionally, the continuous reading of data from disks when scrubbing can result in wear and tear, especially on larger capacity disks, due to the significant time and energy consumption involved. To address these issues, we propose a selective disk scrubbing method that enhances the overall reliability and power efficiency in data centers. Our method employs a Machine Learning model based on Mondrian Conformal prediction to identify specific disks for scrubbing, by proactively predicting the health status of each disk in the storage pool, forecasting n-days in advance, and using an open-source dataset. For disks predicted as non-healthy, we mark them for replacement without further action. For healthy drives, we create a set and quantify their relative health across the entire storage pool based on the predictor's confidence. This enables us to prioritize selective scrubbing for drives with established scrubbing frequency based on the scrub cycle. %By doing so, we optimize energy consumption and reduce the carbon footprint of data centers.
The method we propose provides an efficient and dependable solution for managing enterprise disk drives. By scrubbing just 22.7\% of the total storage disks, we can achieve optimized energy consumption and reduce the carbon footprint of the data center.
\end{abstract}

\section{Introduction}
\label{sec:intro}
A large-scale data center is a complex ecosystem primarily consisting of various types of storage devices, such as hard disk drives (HDD), solid-state drives (SSD), and hybrid storage devices, spread over multiple geographic locations. As the number of storage components grows in the storage ecosystem, it becomes increasingly difficult to manage its business continuity \citep{bajgoric2022downtime} because of the underlying complexity and uncertainty associated with each individual component's internal working mechanics. For instance, a data center can have a mixed workload of transactional databases, network-attached storage (NAS), and archival storage. Each workload has its unique characteristics and impact on the reliability and remaining useful life of the storage system. Inadequate management of these storage components can result in data loss and eventually business disruption \citep{pinheiro2007failure}. Monitoring the health of storage components is a common proactive approach to maintaining the reliability of storage systems. This is typically achieved through system logs and SMART logs of HDDs and SSDs \citep{zhang2023multi}.

In the context of data centers, many device manufacturers and vendors \citep{188450, vishwakarma2021method} utilize drive failure analysis as a metric to assess the overall reliability of the data center. Drive failure analysis typically categorizes failures into two types: complete failure and latent failure of the disk drive. Complete failure is relatively easy to detect and enables a transparent approach for replacing failed disks. However, latent failure is nearly undetectable and can lead to sudden failure and loss of critical data \citep{bairavasundaram2007analysis, schroeder2010understanding}. 

Traditional system-level approaches, such as storage virtualization technology like Redundant Array of Independent Disks (RAID), tend to focus on passive fault tolerance rather than proactive approaches. Alternatively, many statistical \citep{hamerly2001bayesian, singh2023system} and machine learning \citep{10.1145/2465470.2465473, sun2019system} approaches have been explored to enhance the reliability of the storage system. Machine learning approaches, while powerful, face challenges when it comes to making accurate predictions on unseen data, as production data is subject to change over time \citep{lu2018learning}, requiring continuous model updates. One argument to consider is that even a small false positive rate (FPR), such as $0.1\%$, can be a concern when implementing the model in real-world data centers that may have millions of disk drives. Instead of using drive failure analysis for reliability enhancement, an alternative approach could be to leverage disk drive scrubbing \citep{iliadis2008disk, iliadis2011disk} in a more fine-grained manner to identify specific disks that require further attention. 

Disk scrubbing is a process of performing full media pack sweeps across allocated and unallocated disks to detect and rebuild latent medium errors \citep{ryu2009effects}, reducing the chances of bad block media detection during host I/O activity. However, running scrubbing tasks for the entire disk population in an array can significantly increase the load on the data storage system, potentially degrading its performance.  %Furthermore, reading data from the disk for scrubbing may also result in wear and tear on the disk. 
Additionally, if the disk has a larger capacity (e.g., 12TB), it will take a considerable amount of time to complete the operation. 

Our proposed method aims to optimize disk scrubbing by selectively targeting only the disks that require the operation. We use a learning framework based on Mondrian conformal prediction, which is agnostic to the specific machine learning algorithm, to identify specific disks for scrubbing. The method involves forecasting the health of a disk n-days ahead through binary classification. A set of healthy drives is created, and the health status of the entire storage pool is quantified based on the predictor’s confidence. Drives marked as unhealthy are then sent for treatment based on the administrator's decision. The metrics obtained from Mondrian conformal prediction are used to prioritize selective scrubbing for the drives. This proposed method has two main advantages: first, it leverages the underlying drive failure analysis, and second, the quantified output from the forecast engine can be used as input for the disk scrubbing scheduler engine, optimizing the scrubbing process and enhancing reliability in the data center environment. A summary of our contributions is as below: 
\begin{itemize}
    \item Introducing a fine-grained approach for identifying disk drives to be scrubbed by complementing the existing failure analysis engine on the storage sub-system with an algorithm-agnostic framework based on Mondrian conformal prediction. 
    \item Translating the output of the framework, i.e. the confidence of prediction for new data points, into a ranking mechanism that sorts the healthy drives in descending order for further treatment by the decision engine.
    \item Implementing a scrubbing frequency schedule for the entire storage array based on n-step ahead system load prediction using probabilistic weighted fuzzy time series, which is mapped to the scrubbing engine for optimized disk scrubbing. 
\end{itemize}
The method contributes to a proactive approach that provides value for business continuity, such as resource and power savings in data centers during data scrubbing by selectively spinning disks down. This means that only the required disks are scrubbed on a priority basis, while the healthy disks can be scrubbed less frequently or not at all. This optimized approach helps improve operational efficiency and reduces unnecessary disk wear, resulting in potential cost savings and enhanced reliability for the data storage system.

This paper is structured as follows: \autoref{sec:motivation} provides the motivation for our work and outline the design considerations. In \autoref{sec:works}, we present a review of related work in the field. We provide a definition and algorithmic overview of conformal prediction in \autoref{sec:preliminaries}. In \autoref{sec:solution}, we deliver a synopsis  of our proposed solution, followed by the presentation of experimental results in \autoref{sec:experiment}. The usability and interpretation of disk drive health metrics are discussed in \autoref {sec:discussion}, and we conclude with \autoref{sec:conclusion}, summarizing our findings.

\section{Motivation and design goals}
\label{sec:motivation}
In data centers, a significant number of unhealthy drives go undetected due to latent failure attributes, resulting in fail-stop scenarios. One common approach to mitigate such scenarios is disk scrubbing, which consists of verifying disk data through a background scanning process to identify bad sectors. However, this process can consume energy and cause performance degradation depending on the trigger schedule. This scenario raises concerns in the industry, especially as disk capacities increase. We notice a missing link in addressing 'which disk to scrub', 'when to scrub', based on frequency of scrub cycle while minimizing storage array performance impact and also maximizing the reliability. In this paper, we consider the following objectives and design approaches to tackle this challenge :

\begin{itemize}
    %\item \textbf{Selective disk drive scrubbing:} 
    \item \textbf{Which disk to scrub?}
    %Disk scrubbing can be time-consuming, particularly for large drives or those with slow transfer rates. 
    Depending on the specific scrubbing process, it can temporarily degrade the performance of the drive. To ensure that the drive remains fast and responsive, minimizing the frequency of scrubbing is crucial. Instead of performing scrubbing for all disks in the storage array, our approach focuses on selectively scrubbing only the disks that require it, thereby reducing the overall time required to complete the process.
    %\item \textbf{Remaining useful life optimization:} Disk drives experience wear and tear with each data read or write operation, which can accumulate over time and result in drive problems. Minimizing the frequency of scrubbing can help extend the lifespan of the drive by reducing the number of read and write operations. Machine learning methods can be used to predict the remaining lifespan of a disk drive based on factors such as its age, usage history, and error rate. This predictive capability can enable proactive scrubbing to prevent data loss by identifying when a drive is likely to fail.
    %\item \textbf{Energy saving and reduced power consumption:} Disk drive scrubbing requires powering on and spinning the drive, which can increase its power consumption. Therefore, our objective is to minimize the frequency of scrubbing to reduce the overall power usage and carbon footprint of the drive, aligning with the goal of environmental sustainability.
    %\item \textbf{Scheduling scrub frequency:} 
    \item \textbf{When to scrub?} 
    We can optimize the disk drive scrubbing schedule by considering factors such as the workload of the system, the importance of the data on the drive, and the availability of resources. This approach ensures that scrubbing is performed at the most appropriate times, minimizing the impact on the overall system performance.
\end{itemize}

%In this paper, we propose the adoption of Mondrian conformal prediction as a solution for disk scrubbing. Our approach focuses on selecting only the relevant disks and creating an optimized scrubbing frequency cycle, addressing the challenges associated with disk scrubbing.

\section{Related Work}
\label{sec:works}
Storage device reliability has long been a critical concern in the industry, and existing solutions often rely on failure analysis of storage systems. However, traditional methods like accelerated life tests \citep{cho2015acclerated} have not proven to be reliable indicators of actual failure rates in production environments. Recent machine learning-based approaches, such as multivariate time-series \citep{yu2019hard} and time-series classification \citep{ircio2022multivariate}, have focused on improving model accuracy, but often lack deep integration of domain knowledge. Moreover, the multi-modal approach by \citep{lu2020making} using performance metrics (disk-level and server-level) and disk spatial location only focuses on fail-stop scenarios, which may not be helpful in detecting latent failures. A most recent study \citep{lu2023perseus} has addressed this issue by investigating grey failures (fail-slow drives) using a regression model to pinpoint and analyze fail-slow failures at the granularity of individual drives.

\begin{comment}
    To an extent, machine learning-based solutions are implemented in storage array for proactive failure analysis and it becomes difficult to replace the existing code base with a new one. For this reason, even our solution complements the existing failure analysis engine with a few lines of code while also considering the storage domain knowledge. So, rather than identifying which disks may fail in the future, the focus steps back and identifies the impacting reasons by understanding the background processes running on the storage system, i.e, disk scrubbing.
\end{comment}

Another important factor of disk scrubbing is the implementation cost and power consumption. \citep{mi2008enhancing} and \citep{8715169} address performance degradation due to scrubbing and propose assigning a lower priority to the background process during idle time, i.e. when the disk drive is not actively engaged in processing data or performing any other tasks. \citep{liu2010modeling} and \citep{oprea2010clean} propose a method to mitigate power consumption and determine when to scrub in systems with inexpensive data but require designing another method to identify less critical data. Drive space management in case of replacing the failed disk is discussed in \citep{paris2010improving}, along with reducing the need for frequent scrubbing. 
%\citep{6263919} contributes to guaranteeing availability by addressing when and how much to scrub. 
A multilevel scrubbing is proposed in \citep{9218551} using a Long Short-Term Memory (LSTM) model to detect latent sector errors in a binary classification setup. However, using machine learning-based models may treat healthy and relatively less healthy disks the same, leading to unnecessary scrubbing of healthy disks.

To the best of our knowledge, our work is the first to adopt Mondrian conformal prediction for assigning a health score to each individual disk drive and using the metrics to design a scrubbing cycle aligned with the system idle time.

\section{Conformal prediction}
\label{sec:preliminaries}

Conformal prediction \citep{shafer2008tutorial} is a powerful framework in machine learning that allows for prediction with uncertainty. Unlike traditional point prediction methods used in classification tasks, conformal prediction provides a set prediction. This means that instead of outputting a single predicted label, conformal prediction provides a range of possible labels that are likely to be correct, along with confidence and credibility measures in the correctness of these predictions. This is particularly useful in cases where the prediction task may be uncertain or when the model is dealing with previously unseen data, especially when the application is of high risk \citep{luo2022sample}.

Conformal prediction is not limited to classification tasks alone, but it is also valid for regression tasks by providing prediction intervals instead of a single-point prediction. These prediction intervals represent a range of possible values for the target variable, along with a measure of confidence in the correctness of these intervals. This allows for more nuanced and interpretable predictions in regression tasks, where the goal is to estimate a continuous value rather than a discrete class label.

One of the key advantages of conformal prediction is its agnostic nature, which means it can be used with any machine learning algorithm. This flexibility allows for the integration of conformal prediction into various machine learning pipelines without being constrained by the choice of a specific algorithm. This makes conformal prediction a versatile tool that can be applied to various tasks and domains \citep{messoudi2020deep}.

A great resource for learning more about conformal prediction is the book "Algorithmic Learning in a Random World" \citep{Vovk2022-wq}, which provides a detailed overview of the theory and applications of conformal prediction. Additionally, a repository of conformal prediction implementations and resources is maintained on GitHub\footnote{\url{https://github.com/valeman/awesome-conformal-prediction}}\citep{manokhin_valery_2022_6467205}, providing practical tools and examples for applying conformal prediction in various machine learning settings. This makes it easier for researchers and practitioners to implement and experiment with conformal prediction in their own work.

The overall approach of conformal prediction in the inductive setting is presented in \autoref{algo:icp}. 

\begin{algorithm}[ht]
\floatconts
{algo:icp}% label
{\caption{Inductive conformal prediction (ICP)}}
{% contents
\textbf{Input}: Divide the training sequence $Z_{t r}$ into two disjoint subsets; the proper training set $Z_t$ and the calibration set $Z_c=\left\{\left(x_1, y_1\right), \ldots,\left(x_q, y_q\right)\right\}$

%Let $z_1=\left(x_1, y_1\right), \ldots, z_n=\left(x_n, y_n\right) \in \mathbf{Z}$, $x_i \in \mathbf{X}$ and $y_i \in \mathbf{Y}=\left\{C_1, \ldots , C_p\right\}$ \\
\textbf{Task}: Predict $y_{n+1} \in \mathbf{Y}$ for any new object $x_{n+1} \in \mathbf{X}$. \\
\textbf{Algorithm}:
\vspace{-1mm}
\begin{enumerate}
  \item Split the original data set $\mathbf{Z}$ into a proper training set $\mathbf{Z}^{\text {\textit{tr}}}$ with $\left|\mathbf{Z}^{\text {\textit{tr}}}\right|=m$ and a calibration set $\mathbf{Z}^{\text {\textit{cal}}}$ with $\left|\mathbf{Z}^{\text {\textit{cal}}}\right|=n-m=q$. 
  \vspace{-2mm}
  \item Train using $h: \mathbf{X} \rightarrow \mathbf{Y}$ on $\mathbf{Z}^{\text {tr }}$ and obtain the nonconformity measure $f(z)$. The non-conformity measure for $h$ is defined as $f(z)=1-\hat{P}_h[y \mid x]$. 
  \vspace{-2mm}
  \item Apply the non-conformity measure $f(z)$ to each example $z_i$ of $\mathbf{Z}^{\textit{cal}}$ to get the nonconformity scores $\alpha_1, \ldots, \alpha_q$. 
  \vspace{-2mm}
  \item Choose a significance level $\epsilon \in(0,1)$ to get a prediction set with a confidence level of $1-\epsilon$ 
  \vspace{-2mm}
  \item For a new example $x_{n+1}$, compute a non-conformity score for each class $C_k \in \mathbf{Y}$ : $\alpha_{n+1}^{C_k}=f\left(\left(x_{n+1}, y=C_k\right)\right)$ 
  \vspace{-2mm}
  \item For each class $C_k \in \mathbf{Y}$, compute the $p$-value : $
  p_{n+1}^{C_k}=\frac{\left|\left\{i \in 1, \ldots, q: \alpha_{n+1}^{C_k} \leq \alpha_i\right\}\right|}{q}$ 
  \vspace{-2mm}
  \item Build the prediction set: $\Gamma^\epsilon=\left\{C_k \in \mathbf{Y}: p_{n+1}^{C_k}>\epsilon\right\}$
  \end{enumerate}
}
\end{algorithm}

\begin{comment}
    \begin{algorithm}[htbp]
\floatconts
{algo:cp}% label
{\caption{The Conformal Algorithm}}
{% contents
\textbf{Input}: Nonconformity measure $A$, significance level $\varepsilon$, examples $z_1, \ldots, z_{n-1}$, object $x_n$, label $y$ \\
\textbf{Task}: Decide whether to include $y$ in $\Gamma^{\varepsilon}\left(z_1, \ldots, z_{n-1}, x_n\right)$. \\
\textbf{Algorithm}:
\begin{enumerate}
  \item Provisionally set $z_n:=\left(x_n, y\right)$.
  \item For $i=1, \ldots, n$, set $\alpha_i:=A(\langle z_1, \ldots, z_n \rangle  \backslash \supsetneq \langle  z_i \rangle , z_i)$.
  \item Set $p_y:=\dfrac{\#\left\{i=1, \ldots, n \mid \alpha_i \geq \alpha_n\right\}}{n}$.
  \item Include $y$ in $\Gamma^{\varepsilon}\left(z_1, \ldots, z_{n-1}, x_n\right)$ if and only if $p_y>\varepsilon$.
\end{enumerate}
}
\end{algorithm}
\end{comment}

\subsection{Mondrian conformal prediction (MCP)}

Mondrian conformal prediction (MCP) is a variant of the conformal prediction framework that provides a guarantee on a subset of the dataset, or on specific categories of the dataset. This variant is originally established for a classification problem by creating class-conditional or attribute-conditional categories \citep{vovk2003Mondrian}. However, a Mondrian version exists for regressors \citep{bostrom2020mondrian}.

In imbalanced datasets, the minority class (i.e., the class with fewer instances) may lead to biased predictions and inaccurate confidence measures. Mondrian conformal prediction is a powerful tool to handle this issue by maintaining the same error rate for both the majority and minority classes, ensuring that the predictions are not biased toward the majority class.

When calculating the non-conformity scores in MCP, we only consider the scores related to the examples that share the same class as the object $x_{n+1}$, which we are testing hypothetically. Consequently, the $p$-value is substituted in step 6 of \autoref{algo:icp} and calculated as: 
$$
p_{n+1}^{C_k}=\frac{\left|\left\{i \in 1, \ldots, q: y_i=C_k, \alpha_{n+1}^{C_k} \leq \alpha_i\right\}\right|}{\left|\left\{i \in 1, \ldots, q: y_i=C_k\right\}\right|}.
$$
Mondrian conformal prediction has been implemented for various domains in many real-life use cases for academia as well as industry. For instance, \citep{messoudi2021class} apply MCP for tenant debt prediction in real estate, \citep{alvarsson2021predicting} use it for modeling ABC transporters in drug discovery, and \citep{vishwakarma2021system} leverage conformal predictors for detecting persistent storage failure analysis in the enterprise storage domain.

\subsection{Evaluation metrics}
%We use the confidence and credibility obtained from Mondrian conformal predictor as performance metrics. Confidence tells about how confident is the prediction and credibility quantifies the quality of the data points. In our case the data point refers to the SMART attributes and system statistics. We provide a formal definition of confidence and credibility as below: 

To evaluate the performeance of conformal prediction models, several metrics can be employed. In our study, we will focus only on two :

\begin{itemize}
    \item \textbf{Confidence}: reflects the certainty of the model that a prediction is a singleton, or a unique outcome. Confidence is based on the concept of p-values, which are used to assess the probability of obtaining an outcome as extreme as the one observed, assuming that the null hypothesis is true. A higher confidence value suggests that the model is more confident about the accuracy of its prediction and that the predicted label is likely to be correct. Conversely, a lower confidence value implies that there may be alternative labels that are equally likely. This metric is defined as:
$$\text{Confidence} (x) = \sup \{1 - \epsilon : |\Gamma_\epsilon (x)| \leq 1\}. $$
\item \textbf{Credibility} quantifies the likelihood that a sample comes from the training set, as determined by the minimal significance level that would result in an empty prediction region. In other words, credibility is expressed as the largest p-value, which serves as the lower bound for the value of the significance level $\epsilon$ that would result in an empty prediction. A higher credibility value indicates a higher likelihood that the sample is consistent with the training set, while a lower credibility value indicates a higher likelihood of the sample being inconsistent with the training set. This metric is defined as:
$$\text{Credibility} (x) =  \inf \{\epsilon : |\Gamma_\epsilon (x)| = 0\}.$$
\end{itemize}

\section{Mondrian conformal prediction for Disk Scrubbing: our approach}
\label{sec:solution}

\begin{figure*}[ht]
    \centering
    \includegraphics[width = 1\linewidth]{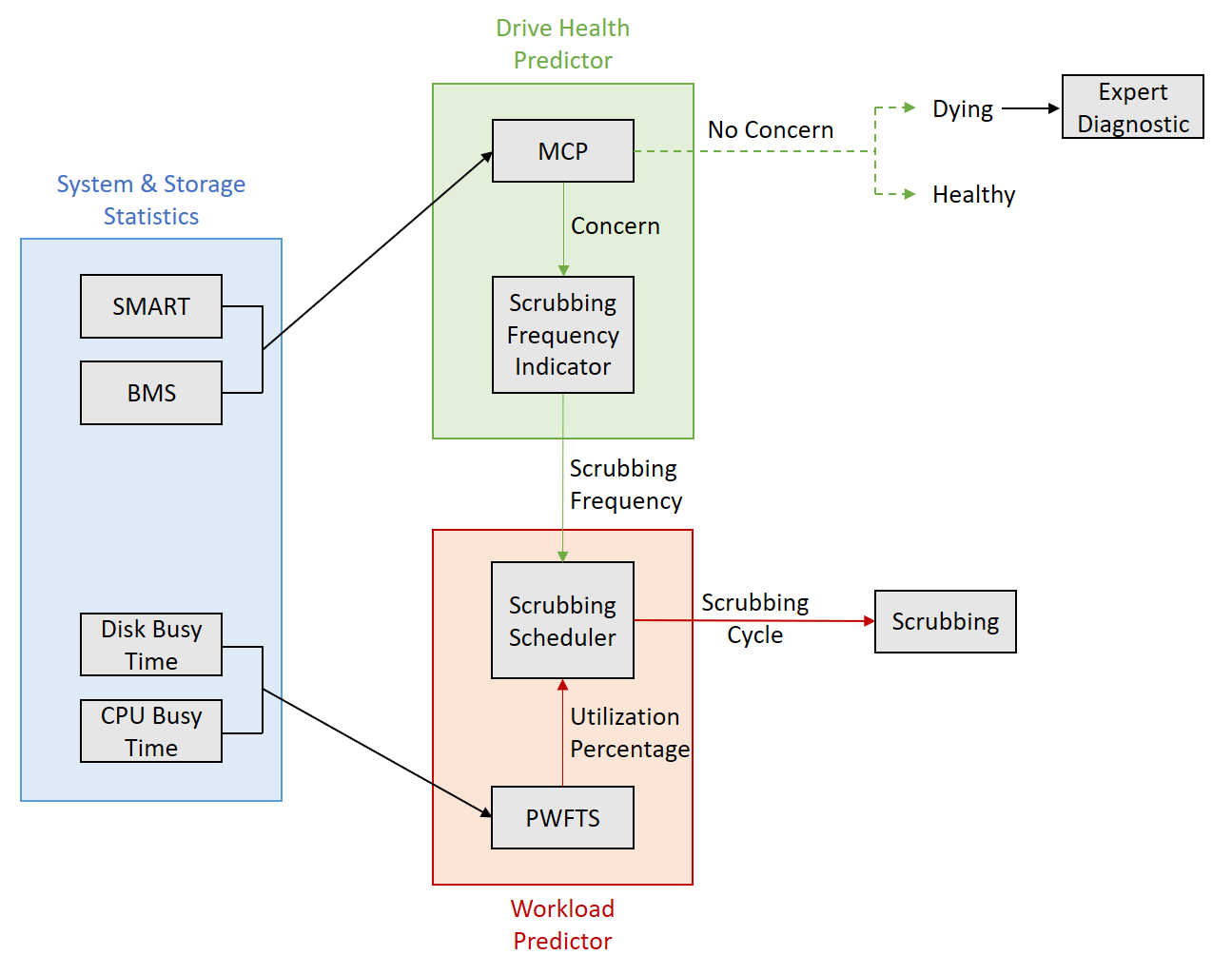}
    \caption{Overall approach of Mondrian conformal disk drive scrubbing.
    }
    \label{fig:architecture}
\end{figure*}

In contrast to the conventional studies mentioned above, we propose a novel approach for disk drive scrubbing based on Mondrian conformal prediction to quantitatively assess the health status of disk drives and use it as a metric for selecting drives for scrubbing. \autoref{fig:architecture} shows a high-level overview of the proposed method.

\begin{comment}
    \begin{figure*}[ht]
    \centering
    \includegraphics[width = 1\linewidth]{figs/framework.png}
    \caption{Overall approach of Mondrian conformal disk drive scrubbing.
    %Optimizing disk scrubbing: a selective approach using a Mondrian conformal predictor's confidence scores for relative health assessment and mapping with scrub frequency.
    }
    \label{fig:architecture}
\end{figure*}
\end{comment}

The proposed architecture consists of three subsystems. The first subsystem is responsible for collecting storage and system statistics, which includes retrieving disk drive data from the storage array, as well as capturing CPU and disk busy statuses. The second subsystem, referred to as the drive health predictor engine, predicts the health status of the drives. It uses MCP to output a set of "No concern" drive disks, i.e. unhealthy/dying drives that can be flagged for manual diagnostics by experts (not discussed in this paper) or completely healthy drives that do not need any scrubbing, as well as a set of "Concern" disks with assigned health scores based on the predictor's confidence, which then are turned into scrubbing frequencies with the scrubbing frequency indicator. The underlying non-conformity score used is margin error function. The third subsystem is the workload predictor engine, which  first predicts the resources' utilization percentage by taking into account SAR logs\footnote{The System Activity Report is a command that provides information about different aspects of system performance. For example, data on CPU usage, memory and paging, interrupts, device workload, network activity, and swap space utilization}, and then combine this result with the scrubbing frequencies in order to schedule when and how frequently disk drive scrubbing is performed. Finally, the scrubbing operation is triggered on the storage array based on the scrubbing cycle. In the following subsections, each component of the overall architecture is described in detail.

\begin{comment}
    In a storage array, there may be a mix of new and old drives, and depending on the workload, certain disks may encounter more reliability issues relative to others in the same array due to their age. Several factors can contribute towards drive aging, including:
\begin{itemize}
    \item \textbf{Bit Rot}: Over time, magnetic charges on a hard drive can degrade, resulting in data corruption. This can pose challenges for scrubbing software to accurately identify and correct errors, as the degradation may affect the effectiveness of the scrubbing process.
    \item \textbf{Aging Software}: As technology evolves, the software used for disk drive scrubbing may become outdated and less effective. In addition, software updates may not be compatible with older drives, which can limit the effectiveness of scrubbing and reduce the ability to address errors.
    \item \textbf{Operating Environment}: The operating environment in which the drives are located can also impact the effectiveness of scrubbing. For instance, high temperatures or humidity levels can accelerate the aging process of drives, causing them to degrade more rapidly. Exposure to magnetic fields or other sources of interference can also introduce errors that are challenging to correct through scrubbing.
\end{itemize}
All of the above factors for low reliability of disk drives can be solved by quantifying the degree of their health as described in \autoref{sec:architecture}
\end{comment}

\subsection{System and Storage statistics}

The main components of this subsystem are:
\begin{itemize}
    \item \textbf{SMART}: stands for Self-Monitoring, Analysis, and Reporting Technology, and refers to a set of predefined parameters provided by device manufacturers that offer insights into various aspects of a storage device's performance, including temperature, error rates, reallocated sectors, and more. Each attribute has a threshold value assigned by the manufacturer, indicating the acceptable limit for that parameter. When a parameter exceeds its threshold value, it may indicate a potential issue with the storage device. We use SMART parameters as input features for the drive health predictor engine.
    
    \item \textbf{BMS}: stands for Background Media Scanning, and
    is a passive process that differs from disk scrubbing, which actively scans the disk for errors during idle periods without reading or writing data. BMS involves scanning the disk for errors in the background without interrupting normal operations. In our proposed architecture, we also extract this BMS feature, which is a numerical value for the number of times it encounters errors while performing a scan on the same drive, and feed it to the drive health predictor engine.

    \item \textbf{Disk and CPU busy time}: The performance of a drive is heavily dependent on its critical processes, such as data access and write speed. The numeric values range between 1 to 100 in terms of percentage and change over time with a sampling period of 1 hour. These system statistics are extracted from the SAR logs (standard logs for system utilization) and converted into time series data, which can then be used by the workload predictor engine. %used for  If a disk takes more time for R/W, then the latency is high and this is used to flag the spike in IO latency.  
\end{itemize}

\begin{comment}
   \subsubsection{Mapping error rates}
Most of the advanced HDD give Data Unit Read and Data Unit Write. This gives us information on the read/write unit performed on a single HDD. Information is easily available on SMART pages. \\
\\
\texttt{
        Data Unit Read: 0x746da4 of 512k bytes.\\
        Data Unit Written: 0x2d0 of 512k bytes.\\
}\\
An additional error log gives us information on errors occurring on an HDD. 
\begin{verbatim*}
error_count:18446744069579341823
sqid:65535
cmdid:0x1
status_field:0x2(INVALID_OPCODE)
parm_err_loc:0xffff
lba:0xffffffff00040003
nsid:0xffffffff
vs:255
\end{verbatim*}
If for read-write numbers more errors are coming, the disk drive will be selected for scrub operation. 
$$ Bytes_{written} + \frac{Bytes_{read}}{rw_{weight}}\geq \frac{1} {BER}.$$ 
\end{comment}

\subsection{Which disk to scrub: Drive health predictor}

In a normal data center setting, all disk drives are classified as either healthy or unhealthy. Unhealthy disks are supposed to be dying or imminently failing, thus they are not marked for scrubbing, while healthy disks are marked for scrubbing. 

In our approach, we propose to assign a relative 'degree of health' score to each disk. Drives that are marked as of \textit{No concern} are either dying/imminently failing or completely healthy, while those marked as of \textit{Concern} have different degrees of health other than failing or healthy. The conformal prediction framework then classifies the "No-concern" and "Concern" drives, and only selects the disks which are in the set of "Concern" drives for further ranking. These are the drives which are concerning to us and is used as input for the scrubbing scheduler.
%The concerned drives are then ranked for selective scrubbing and mapped with the appropriate scrub frequency. 

\begin{figure}[ht]
    \centering
    \includegraphics[width = 0.5\linewidth]{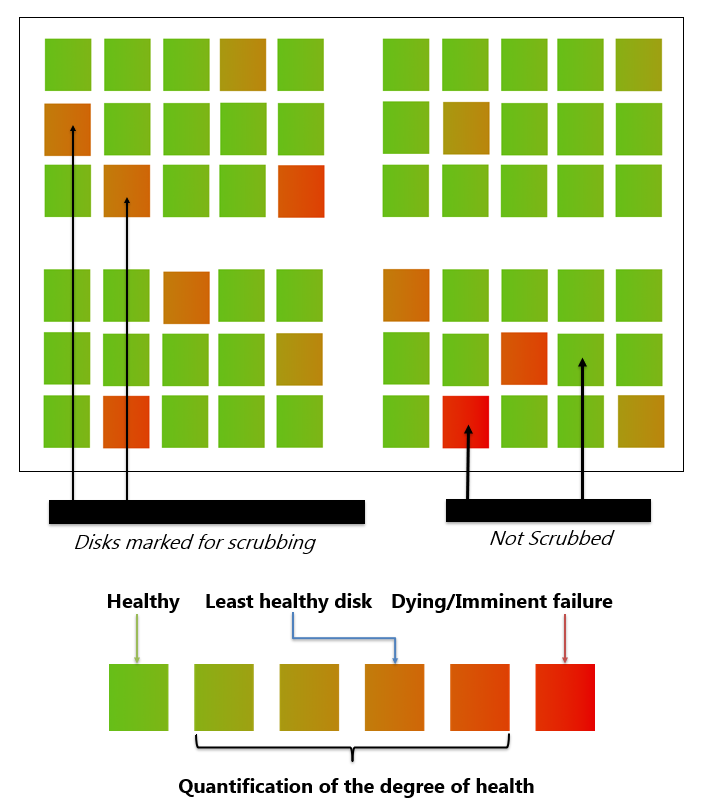}
    \caption{Quantifying the health of disk drives: The disks which are healthy and non-healthy are not selected for scrubbing, while the disks of concern are marked for scrubbing.}
    \label{fig:proposedmethod}
\end{figure}

Our focus, as shown in \autoref{fig:proposedmethod}, is on identifying disks in the system that are currently of concern or may become concerning soon, and only selecting those disks for scrubbing. This approach reduces the number of disks meant for scrubbing, since even completely healthy drives are not scrubbed, making the process more efficient and targeted. By doing so, we optimize time, power, and energy consumption and reduce the carbon footprint of data centers.

When dealing with disk drives in a usual data center environment, failures are rare over a period of time, resulting in a highly imbalanced dataset with a small number of failed disks and the majority of disks being healthy. To handle this imbalanced data, we adopt a Mondrian Conformal Prediction approach, in order to get the prediction labels "0": failed and "1": healthy, along with their confidence score that serves as a health score in our case. This means that our MCP algorithm selects disks with a confidence score depending on the threshold chosen by the administrator.

For instance, if the administrator sets a threshold of $1\%$, this will lead to excluding disks with health scores above $99\%$ as healthy or failing (depending on the label) and only selecting disks with a health score lower than $99\%$ for scrubbing. Furthermore, the selected drives can be mapped to distinct scrubbing frequencies. Thus, drives with poor health scores may require more frequent scrubbing (every week), while those with good health scores will need less frequent scrubbing (every $3$ months). For the same threshold of $1\%$, the administrator can then map the disk health with a scrubbing frequency, as in \autoref{tab:mapping}.

\begin{table}[ht]
\begin{center}
\begin{tabular}{lll}
 \toprule
\textbf{Disk health} & \textbf{Scrubbing frequency} & Health score \\ \midrule
Best                 & LOW (CYCLE – A)          & $[95\%, 99[$    \\
Medium               & MEDIUM (CYCLE - B)       & $[80\%, 95[$    \\
Poor                 & HIGH (CYCLE – C)         & $< 80\%$   \\ \bottomrule
\end{tabular}
\caption{Mapping of the disk health with the scrubbing frequency based on health score.}
\label{tab:mapping}
\end{center}
\end{table}

\subsection{When to scrub: Workload predictor}
After identifying the disks to be scrubbed using the drive health predictor engine, the next step is to determine the optimal time to perform scrubbing using the workload predictor. This component needs to consider the availability of system resources, i.e. disk and CPU utilization information in the system and storage statistics subsystem.

\begin{figure}[ht]
    \centering
    \includegraphics[width = 0.9\linewidth]{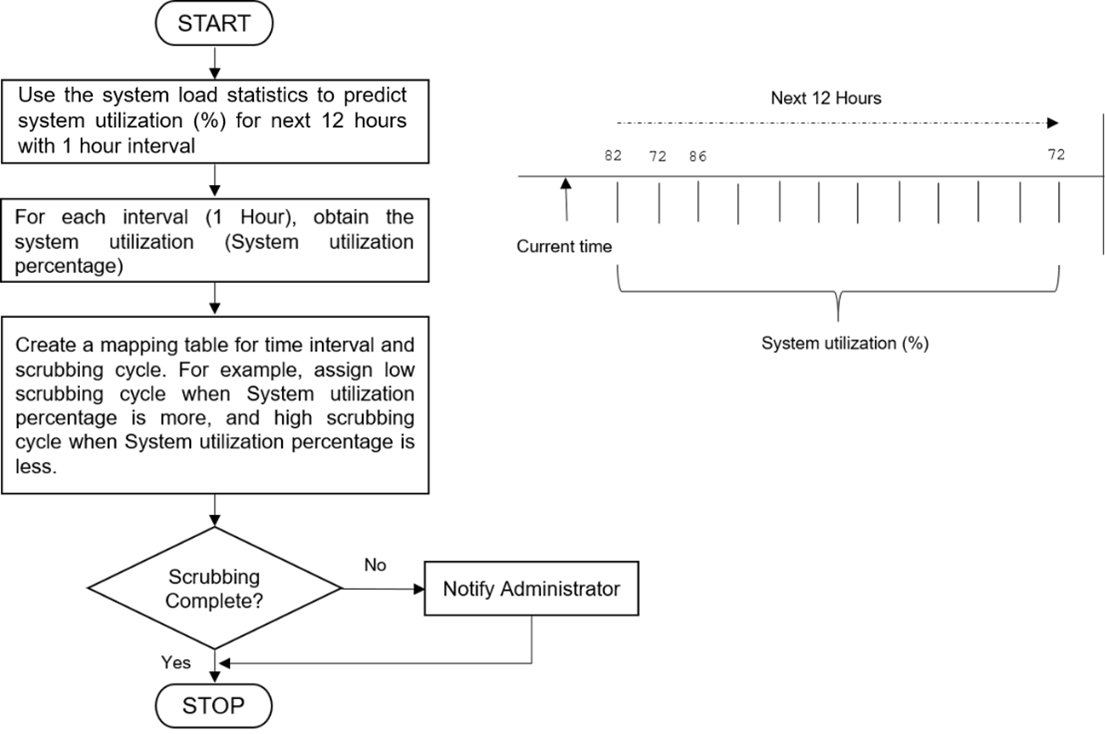}
    \caption{Flowchart of the workload predictor using the PWFTS algorithm.}
    \label{fig:wp}
\end{figure}

The workload predictor employs a Probabilistically Weighted Fuzzy Time Series algorithm (PWFTS), as detailed in \citep{orang2020solar}. This algorithm forecasts n-days ahead system utilization, by predicting the system utilization percentage for the next 12 hours, with 1-hour intervals. Then, this information is combined with one of the three possible scrubbing cycles (A, B, or C as in \autoref{tab:mapping})  obtained from the drive health predictor. Finally, the scrubbing is triggered. During the 1-hour interval, if the scrubbing is complete, then we stop, if not, the administrator is notified. The high-level flowchart for the system workload predictor is outlined in \autoref{fig:wp}.

\iffalse 

\begin{algorithm}[ht]
\floatconts
{algo:workloadpredictor}% label
{\caption{Workload predictor}}
{% contents
\textbf{Input}: SAR logs, scrubbing\_frequency. \\
\textbf{Task}: Predicted system utilization percentage. \\
\textbf{Algorithm}: \\
\vspace{-3mm}

system\_load\_stats $\leftarrow$ GetSystemLoadStatsFromSAR()\\ \vspace{-3mm}

predicted\_utilization $\leftarrow$ PWFTS(system\_load\_stats, 12, 1)

\While{True}{
utilization\_percentage $\leftarrow$ GetSystemUtilizationPercentage()\\
scrubbing\_cycle $\leftarrow$ MapUtilizationToScrubbingCycle(utilization\_percentage, scrubbing\_frequency)
}
\iffalse
\Function{ProbabilisticallyWeightedFuzzyTimeSeries(data, n, step)}{\\
    \KwResult{predicted utilization percentage}
}

\Function{GetSystemLoadStatsFromSAR}{ }{\\
   \KwResult{system load statistics} 
}

\Function{GetSystemUtilizationPercentage}{ }{ \\
    \KwResult{system utilization percentage}}

\Function{MapUtilizationToScrubbingCycle}{utilization\_percentage}{\\
    \KwResult{scrubbing cycle}}
}

\hline
\end{algorithm}

\fi

\iffasle 

The algorithm starts by retrieving disk and CPU busy time data through the GetSystemLoadStatsFromSAR() function, saving the output in the system\_load\_stats variable. Subsequently, the PWFTS function is used to predict the system utilization percentage for the next 12 hours, with 1-hour intervals, and the predicted values are stored in the predicted\_utilization variable. The algorithm then enters a loop that executes every hour, where it retrieves the current system utilization percentage using the GetSystemUtilizationPercentage() function and assigns it to the utilization\_percentage variable. The utilization percentage is then mapped to one of the three possible scrubbing cycles (A, B or C as in \autoref{tab:mapping}) using the MapUtilizationToScrubbingCycle() function, with the resulting output assigned to the scrubbing\_cycle variable. Finally, the algorithm triggers a scrubbing operation based on the scrubbing schedule window and employs the scrubbing\_cycle variable as the input parameter.
\fi 

\begin{figure}[ht]
    \centering
    \includegraphics[width = 0.78\linewidth]{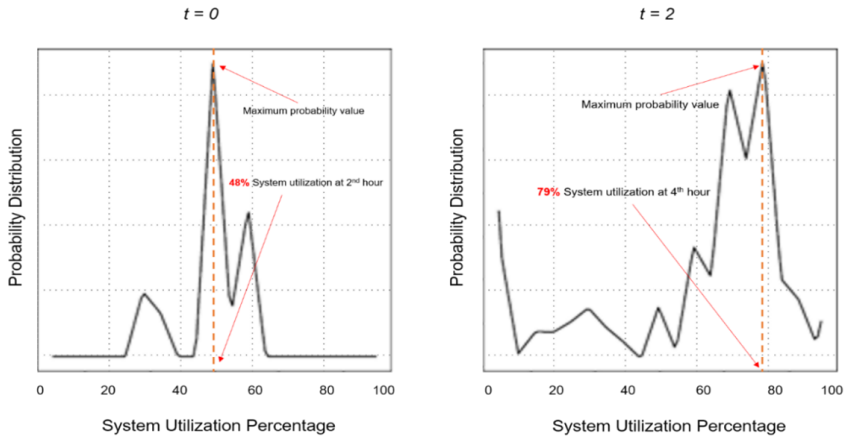}
    \caption{Probability distribution of system utilization percentage for n-days ahead forecasting.}
    \label{fig:predict_interval}
\end{figure}

In \autoref{fig:predict_interval}, we showcase the n-days ahead forecasting of the system utilization percentage. It is evident from the figure that the system exhibits a lower load on day 0 and a higher load on day 2. Consequently, scheduling the scrubbing operations at day 0, when the system is under a lower load, would be more favorable. This approach optimizes the utilization of system resources, ensuring efficient scrubbing of the disks, and leading to lower processing time, lower energy consumption, and a reduced carbon footprint of the data center.

\section{Experimental setting}
\label{sec:experiment}
In this section, we detail the dataset used for our study and the conducted experiments as well as their results.

\subsection{Open-source Baidu dataset}

This dataset \citep{drtycoon_2023} consists of samples collected from Seagate ST31000524NS enterprise-level HDDs, with a total of $23395$ units and $13$ features describing SMART attributes as shown in \autoref{table:smart-ids}. The labeling of each disk was based on its operational status, categorized as either functional or failed. A significant proportion of disks, totaling $22962$, were classified as functional, while a smaller subset of $433$ was marked as failed, resulting in an imbalanced dataset. The SMART attribute values were recorded at an hourly interval for each disk, generating $168$ samples per week for operational disks which gives 1,048,573 actual rows in the dataset corresponding to 23,395 disks (sampling frequency of 1 hour over a period of 2 years). The number of rows represents only the sample of operational disks that are provided in the dataset. However, the failed disks had varying numbers of samples, up to $20$ days prior to failure.

\begin{table}[ht] 
\begin{center}
\begin{tabular}{lll}
\toprule
 Column No             & \textbf{Feature}                         & \textbf{Description}          \\ \midrule
1          & Index of the disk serial number          & Ranging from 1 to 23395       \\ 
2          & Value of SMART ID \#1                    & Raw Read Error Rate           \\
3          & Value of SMART ID \#3                    & Spin Up Time                  \\ 
4          & Value of SMART ID \#5                    & Reallocated Sectors Count     \\
5          & Value of SMART ID \#7                    & Seek Error Rate               \\ 
6          & Value of SMART ID \#9                    & Power On Hours                \\ 
7          & Value of SMART ID \#187                  & Reported Uncorrectable Errors \\ 
8          & Value of SMART ID \#189                  & High Fly Writes               \\
9          & Value of SMART ID \#194                  & Temperature Celsius           \\
10         & Value of SMART ID \#195                  & Hardware ECC Recovered        \\
11         & Value of SMART ID \#197                  & Current Pending Sector Count  \\
12         & Raw Value of SMART ID \#5                & Reallocated Sectors Count     \\
13         & Raw Value of SMART ID \#197              & Current Pending Sector Count  \\
14         & Class label of the disk                  & 0 for failed and 1 for functional \\ \bottomrule

\end{tabular}
\caption{Features' description for the Open-source Baidu dataset.}
\label{table:smart-ids}
\end{center}
\end{table}

\subsection{Experimental results}

For our experiments, we employed the Python programming language and used the MAPIE\footnote{\url{https://github.com/adamzenith/MAPIE/tree/Mondrian}} library \citep{mapieMAPIEModel} for implementing Mondrian Conformal Prediction. The underlying algorithm employed in our experiments was the k Nearest Neighbors (kNN) algorithm.

The main goal of conducting the experimental evaluation is to showcase the significant reduction in the number of disk drives to be scrubbed that can be achieved by using the drive health predictor engine, i.e. exploiting the Mondrian conformal predictor. 

\autoref{table:cm} shows a comparison between the confusion matrix for the drive disk classification problem using the underlying algorithm alone kNN and adding Mondrian Conformal Prediction, where label "0" indicates a disk failure and label "1" indicates a functional one. We can notice that, adding MCP, the number of disks correctly classified as failing has increased from $51314$ to $51669$, i.e., a difference of $355$. This shows MCP helps to identify more disks of the minority class, but with a drawback that is a decrease in the number of disks correctly classified as healthy which has reduced from $296689$ to $268616$, i.e., a difference of $28073$.

\begin{table}[ht]
\begin{center}
\begin{tabular}{lcllll}
\hline
       &           & \multicolumn{2}{c}{kNN} & \multicolumn{2}{c}{MCP} \\ \hline
       & Predicted & 0          & 1          & 0          & 1          \\ \hline
Actual & 0         & 51314      & 547        & 51669      & 537        \\ \hline
       & 1         & 975        & 296689     & 28703      & 268616     \\ \hline
\end{tabular}
\caption{Comparison of confusion matrix results for disk drive classification using kNN and MCP.}
\label{table:cm}
\end{center}
\end{table}

\begin{figure}[ht]
    \centering
    \includegraphics[width = 0.9\linewidth]{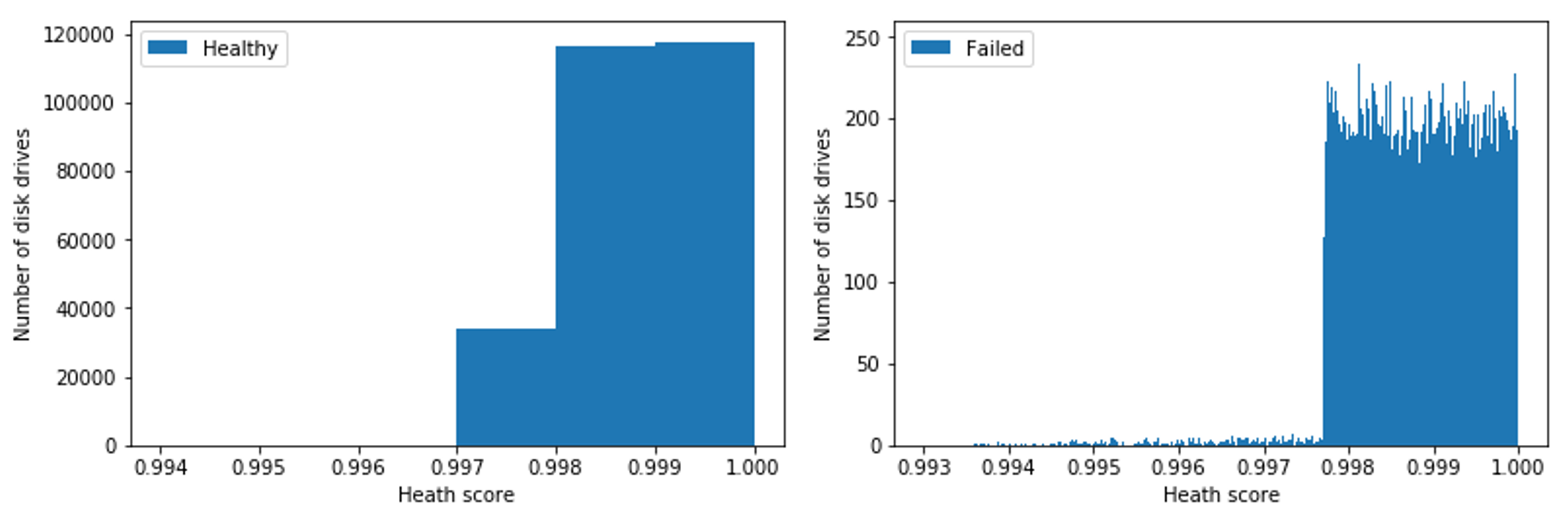}
    \caption{Histogram of the health scores corresponding to healthy predictions (left) and faulting predictions (right).}
    \label{fig:health}
\end{figure}

This issue can be solved by considering the confidence scores and their respective health status, as shown in \autoref{fig:health}. There are nearly 126,224 drives with a health score greater than 99.95\% for the disks labeled as healthy (left), out of total 349,525 disks, but when considering the relative health score, we categorize the 79,396 disk drives with a health score less than 99.9\% as less healthy. Consequently, as shown in \autoref{table:count}, we only select these 79,396 disk drives for scrubbing and skip the remaining 270,129. This approach significantly reduces the number of disks to be scrubbed to only 22.7\%, resulting in lower power and energy consumption, which is noteworthy. 

\begin{table}[]
\begin{center}
\begin{tabular}{lllll}
\hline
Heath score             & 0.998 - 0.9985 & 0.9985 - 0.999 & 0.999 - 0.9995 & 0.9995 and 1 \\ \hline
Disk count  & 16468          & 62928          & 63087          & 126244       \\ \hline
\end{tabular}
\caption{The number of relatively healthy drives based on the health score intervals}
\label{table:count}
\end{center}
\end{table}

\section{Discussion}
\label{sec:discussion}
The proposed method for disk identification for scrubbing offers a dual benefit. Firstly, it can be utilized to assess the reliability of the storage system. Secondly, it employs a disk ranking mechanism to assign relative health scores to individual disks. The choice of classification algorithm depends on factors such as dataset size and available compute resources. However, the decision can be guided by the expertise of the system administrator.

In addition, we discuss how the use of the Mondrian conformal predictor can aid in identifying latent failures of disks, which could be a potential area for future research. Furthermore, we identify three key aspects for designing optimal scheduling and cover performance metrics, including effective coverage and size of the average prediction set.

Lastly, we provide a hypothetical evaluation of energy and power savings resulting from selective scrubbing. This showcases the potential benefits of the proposed method in terms of reduced power and energy consumption, highlighting its effectiveness in optimizing disk scrubbing operations.

\subsection{Optimal scheduling aspect}

With respect to disk scrubbing frequency scheduling, we can design three aspects of scheduling: time window, frequency, and space allocation. Each of them is described below:

\begin{itemize}
    \item Time window focuses on scheduling the time window for scrubbing based on the workload pattern. Scrubbing is done when the system is predicted to be idle.

    \item Frequency involves scheduling the frequency of scrubbing based on the health status of the drive. For drives with the best health, scrubbing is done less frequently. For drives with medium health, scrubbing is done more frequently.

    \item Space deals with scheduling space allocation based on the spatial and temporal locality of sector errors. Instant scrubbing is performed on problematic chunks to ensure efficient disk scrubbing.
\end{itemize}

\subsection{Performance metrics}

\begin{figure}[ht]
    \centering
    \includegraphics[width = 0.9\linewidth]{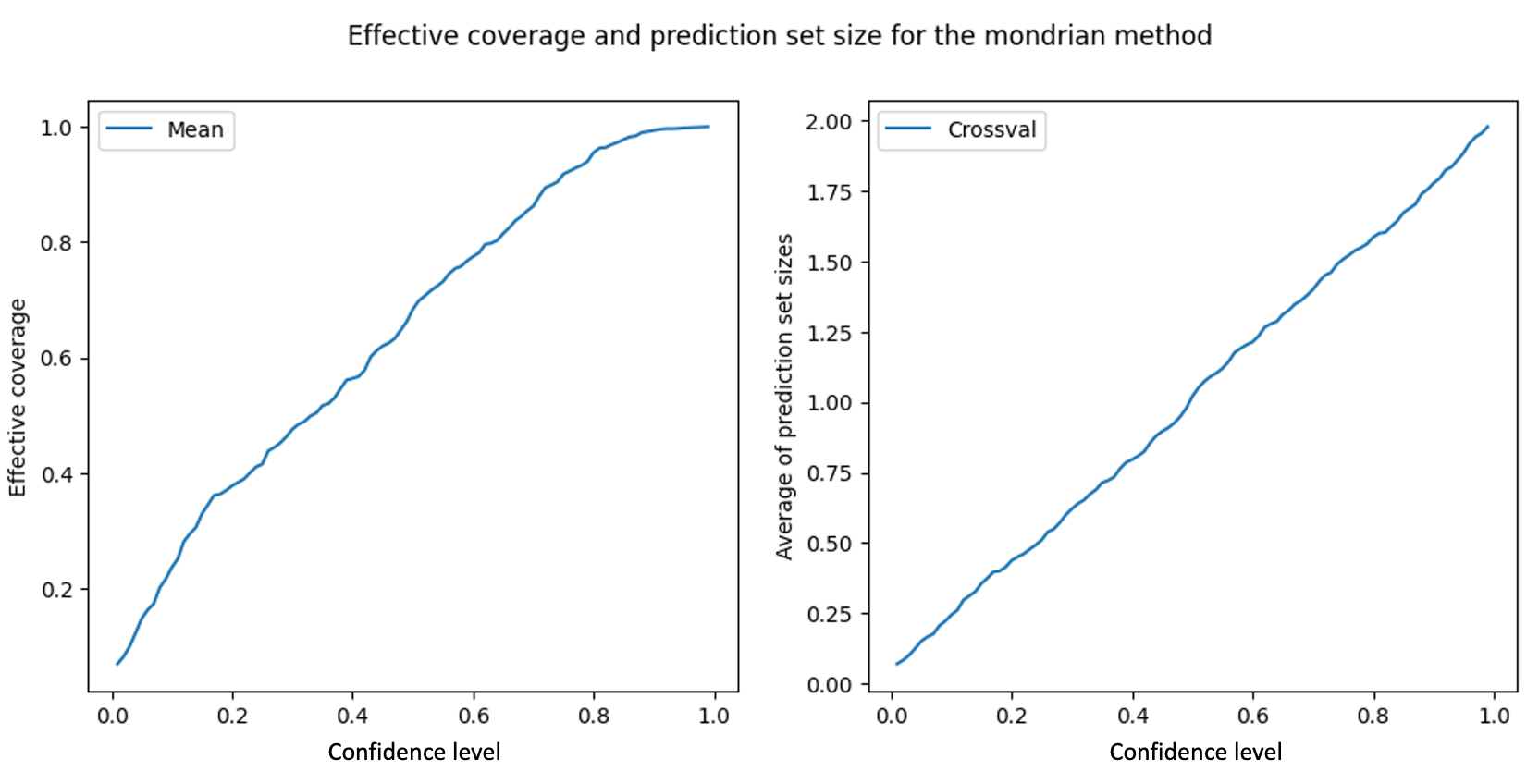}
    \caption{Effective coverage and average prediction set size plotted against confidence level}
    \label{fig:coverage}
\end{figure}

We captured the effective coverage (i.e., for any chosen confidence level, prediction intervals will fail to include the correct label) and prediction set size for the open source dataset in \label{sub:opensource}. The plot in \autoref{fig:coverage} demonstrates that there is a positive correlation between the confidence level and the coverage. The split-conformal method results in a higher mean coverage than the cross-validation method, indicating that the calibration set selection has a considerable influence on the effective coverage. Furthermore, the right side of the figure displays the average size of the prediction set, which increases as the confidence level increases. Similarly, the split-conformal method yields a consistently higher mean prediction set size than the cross-validation method. The metrics can be used to evaluate how well the Mondrian conformal predictor is performing.

\subsection{Power saving from selective scrubbing}

Scrubbing is a resource-intensive operation that can impact the performance of the system during its execution. The time taken to complete a scrubbing operation depends on various factors, such as the size of the HDD being scrubbed. For instance, scrubbing a 1TB HDD may take a few to several hours, while scrubbing an 8TB HDD could take significantly longer, potentially a day or more. Assuming an average power consumption of 7 watts during a 6-hour scrubbing operation for a single HDD, the total energy consumed would be 42 watt-hours (Wh). It's important to note that power consumption during scrubbing can vary for different disks in a data center, depending on factors like disk size, manufacturer, and storage operations. Taking an average value for power usage comparison, if selective scrubbing is performed on 28,000 disks instead of scrubbing all 120,000 disks in a data center based on results from the Baidu open-source dataset, significant power and energy savings can be achieved for the entire data center. 

\section{Conclusion}
\label{sec:conclusion}
The complexity and uncertainty of individual storage components in large-scale data centers pose challenges to business continuity. While proactive approaches like monitoring and failure analysis have been implemented, machine learning approaches may have false positive concerns in real-world applications with numerous disk drives. In this paper, we propose a fine-grained approach to disk scrubbing using a learning framework based on Mondrian conformal prediction, evaluated on the Baidu open-source dataset.

Our method provides a modest yet effective contribution from a methodological perspective. It tackles the issue of aggressive scrubbing of the entire storage array by utilizing Mondrian conformal predictors to assign health scores to each drive and selectively targeting disks with lower scores for scrubbing. This approach generates a prioritized list for the scheduler engine, leveraging drive failure analysis and quantifying disk health across the entire storage pool. As a result, only 22.7\% of the drives need to be scrubbed, leading to power savings and improved reliability.

Future work could involve incorporating Venn-Abers predictors, which offer calibrated probabilities for predictions and could further enhance the accuracy and effectiveness of our approach \citep{vovk2012venn}. By incorporating such predictors, we could potentially refine and optimize our method for even better performance in identifying and addressing potential disk failures in large-scale data centers.

\iffalse 

These commands have optional arguments and have a starred
version. See the \textsf{natbib} documentation for further
details.\footnote{Either \texttt{texdoc natbib} or
\url{http://www.ctan.org/pkg/natbib}}

The bibliography is displayed using \verb|\bibliography|.

\acks{Acknowledgements go here.}
\fi

\bibliography{pmlr-copa23}

\end{document}